\documentclass[aps,prb,floatfix,twocolumn,showpacs]{revtex4}
\usepackage{amsmath}
\usepackage{graphicx}
\usepackage{amsfonts}
\usepackage{amssymb}

\begin{document}

\title{Conserved spin and orbital phase along carbon nanotubes connected
with multiple ferromagnetic contacts}

\author{C. Feuillet-Palma$^{1,2}$, T. Delattre$^{1,2}$, P. Morfin$^{1,2}$, J.-M. Berroir$^{1,2}$, \\
G. F\`eve$^{1,2}$, D.C. Glattli$^{1,2,3}$,B. Pla\c cais$^{1,2}$, A.
Cottet$^{1,2}$ and T. Kontos$^{1,2}$\footnote{To whom correspondence
should be addressed: kontos@lpa.ens.fr}}
\affiliation{$^{1}$Ecole
Normale Sup\'erieure, Laboratoire Pierre Aigrain, 24, rue Lhomond,
75231 Paris Cedex 05,
France\\ $^{2}$CNRS UMR 8551, Laboratoire associ\'e aux universit\'es Pierre et Marie Curie et Denis Diderot, France\\
$^{3}$Service de physique de l'\'etat Condens\'e, CEA, 91192
Gif-sur-Yvette, France.}

\pacs{73.23.-b,73.63.Fg}

\begin{abstract}
We report on spin dependent transport measurements in carbon
nanotubes based multi-terminal circuits. We observe a
gate-controlled spin signal in non-local voltages and an anomalous
conductance spin signal, which reveal that both the spin and the
orbital phase can be conserved along carbon nanotubes with multiple
ferromagnetic contacts. This paves the way for spintronics devices
exploiting both these quantum mechanical degrees of freedom on the
same footing.
\end{abstract}

\date{\today}
\maketitle

\section{Introduction}

The scattering imbalance between up and down spins at the interface
between a non-magnetic metal and a ferromagnetic metal is at the
heart of the principle of the magnetic tunnel junctions or
multilayers celebrated in the field of spintronics
\cite{Baibich:88,Binasch:89}. Although these devices use the quantum
mechanical spin degree of freedom and electron tunneling, they do
not exploit a crucial degree of freedom involved in quantum
mechanics: the phase of the electronic wave function. In most of the
devices studied so far, this aspect has not been developed owing to
the classical-like motion of charge carriers in the conductors used
\cite{Zutic:04}.

Quantum wires or molecules have emerged recently as a promising
means to convey spin information
\cite{Patsupathy:04,Sahoo:05,Man:06,Hauptman:08,Hueso:07,SST:06,Gunnar:08,Tombros:06}.
In these systems, the electronic gas is confined in two or three
directions in space, making quantum effects \textit{a priori}
prominent. In this context, most of the studies have been carried
out in two terminal devices, i.e. with two ferromagnetic contacts.
The need for integration and more complex architectures for
manipulating spin information
\cite{Brataas:00,Huertas:00,Cottet:04,Adagideli:06,Koenig:07} brings
on the question of what happens when a spin active nanoscale
conductor is connected to more than two reservoirs. Multi-terminal
transport has been central in (spin independent) mesoscopic physics,
in particular with the observation of non-local electric signals due
to the delocalization of electronic wave functions
\cite{Buttiker:86,Webb:89,Makarovski:07}. Can this quantum
mechanical non-locality survive and ultimately be exploited in
spintronics devices combining nanoscale conductors and ferromagnets
?

\begin{figure}[!hpth]
\centering\includegraphics[height=0.95\linewidth,angle=0]{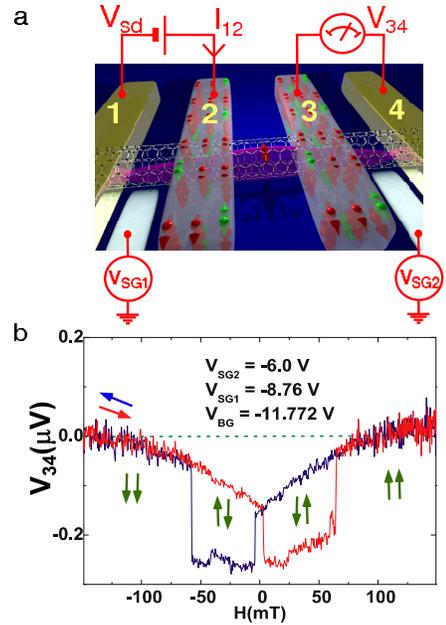}
\caption{\textbf{a.} Schematics diagram of the devices studied in
the article. \textbf{b.} Non-local voltage $V_{34}$ for sample I as
a function of the external magnetic field $H$ for side gate voltages
$V_{SG1}=-8.76V$, $V_{SG2}=-6.00V$ and a back gate voltage
$V_{BG}=-11.772V$.} \label{device}
\end{figure}

In this article, we address this question through multi-terminal
spin dependent transport measurements in single wall carbon
nanotubes (SWNTs) with ferromagnetic and non-magnetic contacts.
Non-local voltage and conductance measurements reveal that the spin
as well as the orbital phase are conserved along the whole active
part of our SWNTs. We observe a \textit{non-local spin field effect
transistor}-like action which is a natural consequence of quantum
interference in a few channel conductor. In spite of the inherent
complexity of the spectrum of our devices, we can account well for
our findings using a simple theory based on a scattering approach.
These results bridge between mesoscopic physics and spintronics.
They open an avenue for nanospintronics devices exploiting both the
spin and the orbital phase degrees of freedom, which could provide
new means to manipulate the electronic spin, because the orbital
phase of the carriers can easily be coupled to the local electric
field in nanoscale conductors.

The principle of non-local transport measurements is to use a
multi-terminal structure with two terminals playing the roles of
source and drain, and the others the role of non-local voltage
probes. Since the pioneering work by Johnson and Silsbee in
metals\cite{Johnson:85}, non-local spin dependent voltages have been
studied in various multichannel diffusive circuits based on
semiconducting heterostructures\cite{Lou:07}, metallic
islands\cite{Zaffalon:05} and graphene\cite{Tombros:05}. These
signals are well captured using a \textit{classical} bipartite
resistors network, with two branches corresponding to opposite spin
directions \cite{ValetFert:93}. The non-local spin signal stems from
the imbalance between the up spin and down spin branches of the
network, which reflects the imbalances between e.g. the two spin
populations. Importantly, this interpretation is valid only when one
can neglect quantum mechanical non-local signals which arise from
the delocalization of the carrier's wave function.

Coherence effects induce only corrections to transport  at low
temperatures in metals or semiconducting heterostructures which
involve many conducting channels\cite{Vila:07}. In contrast,
coherence becomes essential in understanding transport in molecules
or quantum wires where quantum mechanics primarily controls
conduction. The studies of non-local spin transport in the coherent
regime have been elusive so far. Here, we use the high versatility
of Single Wall carbon NanoTubes (SWNT) to achieve the required
devices and to explore these phenomena. We observe a gate-controlled
spin signal in non-local voltages and an anomalous spin conductance
which are specific to the coherent regime.

\section{Experimental setup}

\begin{figure}[!pth]
\centering\includegraphics[height=0.95\linewidth,angle=0]{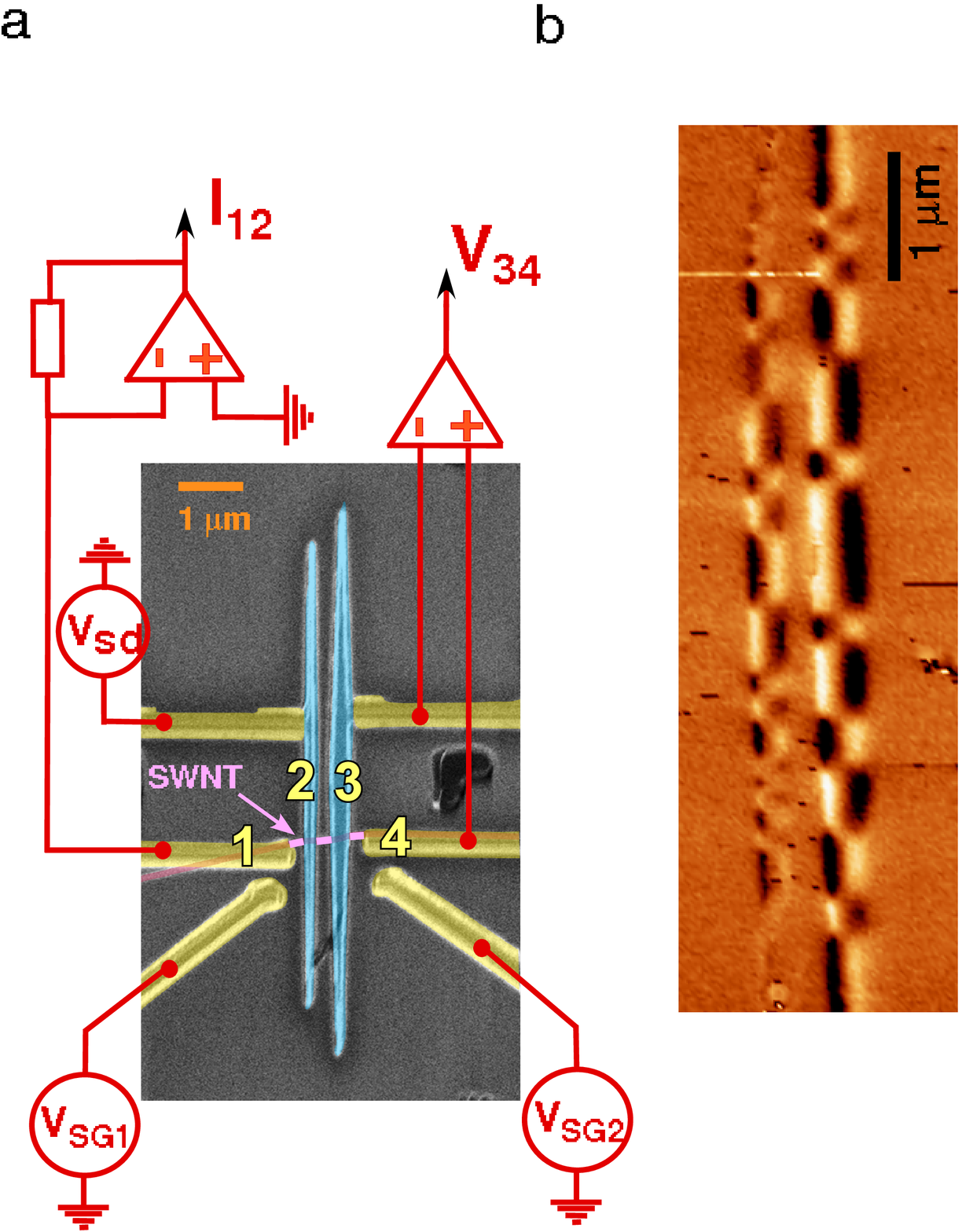}
\caption{\newline \textbf{a.} SEM picture of sample I. The NiPd
electrodes are highlighted in blue and the Pd stripes are
highlighted in yellow. The SWNT is highlighted in purple. The orange
scale bar is $1\mu m$. \textbf{b.} MFM characterization of the NiPd
electrodes at room temperature on a test device similar to sample I
without the SWNT. The black scale bar is $1\mu m$.
}%
\label{deviceSEM}%
\end{figure}

We use the measurement scheme represented on figure \ref{device}a.
Our devices are made out of a SWNT connected to 4 electrodes
labelled 1,2,3 and 4 from the left to the right, with 2 and 3
ferromagnetic $NiPd$ electrodes and 1 and 4 non-magnetic $Pd$
electrodes.  In addition, the device is capacitively coupled to a
back-gate electrode with voltage $V_{BG}$ and two side gate
electrodes with respective voltages $V_{SG1}$ and $V_{SG2}$, acting
mainly on sections $12$ and $34$ of the nanotube respectively.
Throughout the paper, the temperature is set to $4.2K$.

 A SEM picture of a typical device is shown in figure
\ref{deviceSEM}a. We use chemical vapor deposition with a standard
methane process to fabricate our SWNTs on a Si substrate. We
localize the SWNTs with respect to Au alignment markers by Scanning
Electron Microscopy (SEM) or Atomic Force Microscopy (AFM). We
fabricate the contacts and gates using standard e-beam lithography
and thin film deposition techniques. We deposit the normal and the
ferromagnetic contacts in one fabrication step using shadow
evaporation techniques. The central ferromagnetic electrodes consist
of a 30nm-thick $Ni_{0.75}Pd_{0.25}$ layer below a 70nm-thick $Pd$
layer. The normal contacts consist of 70nm of Pd. Such a method
allows to achieve two probe resistances as low as 30kOhm between the
normal and the ferromagnetic reservoirs. In addition to the highly
doped Si substrate with $500nm$ $SiO_{2}$ which is used as a global
backgate, we fabricate two side gates whose voltages $V_{SG1}$ and
$V_{SG2}$ are used to modulate transport in our devices. Each
nanotube section defined in this manner has a length ranging from
$300nm$ to about $600nm$.

Our measurements are carried out applying an AC bias voltage
$V_{sd}$ of about $200$ to $300 \mu V$ between the normal electrode
$1$ and the ferromagnetic electrode $2$, at a typical frequency of
$77.7Hz$. This generates a finite non local voltage $V_{34}$ between
the ferromagnetic electrode $3$ and the normal electrode $4$. We
also measure simultaneously the conductance $G=dI_{12}/dV_{sd}$.
Note that a finite $V_{34}$ has already been observed in similar but
non-magnetic devices due to coherent propagation of electrons in the
SWNT and lifting of the K/K' degeneracy\cite{Makarovski:07}. Here,
we focus on the specific effects due to ferromagnetic leads. A spin
contrast is obtained by comparing the electric signals in the
parallel (P) configuration (magnetizations of electrodes 2 and 3
pointing in the same direction) and in the (AP) configuration
(magnetizations pointing in opposite directions). A finite magnetic
field is applied in plane parallel to the easy axis of the
ferromagnetic electrodes (for samples I and III), which is
transverse as shown by MFM characterization carried out at room
temperature (see figure \ref{deviceSEM}b). The observed magnetic
contrast shows the presence of large transverse domains of a typical
size of $1 \mu m$. Due to the different widths of respectively
$150nm$ and $250nm$, the coercive fields of the two ferromagnetic
electrodes are different. Generally, this leads to a sharp switching
at about $50mT$ for one of the electrodes. For the lower field
switching, it turns out to be more difficult to obtain
systematically switchings as sharp as those of sample I. The P and
AP configuration can be obtained selectively by sweeping the
external magnetic field. We determine $MV=V^{P}_{34}-V^{AP}_{34}$
and $MG=100 (1-G^{AP}/G^{P})$.

\section{Gate controlled non-local spin signal.}

The magnetic field dependence of the non-local voltage $V_{34}$ of
sample I is shown in figure \ref{device}b. Upon increase and
decrease of the external magnetic field $H$, the characteristic
hysteretic switching of a spin valve is observed. We observe sharp
switchings which show that the external field is well aligned with
the magnetic anisotropy of both electrodes in this case. Upon
increasing $H$ (see red line in figure 1b), we obtain the AP
configuration for $H \in [10mT ; 50mT]$, and the P configuration
otherwise. For the particular gate voltage set used in figure
\ref{device}b, $V_{34}$ changes from $V^{P}_{34}=-0.12 \mu V$ to
$V^{AP}_{34}=-0.25 \mu V$ upon switching from the P to the AP
configuration, leading to a finite $MV$. Unlike the majority of our
samples, this spin signal is superimposed to an intrinsic background
here (see discussion in section IV). A finite $MV$ has already
reported in multichannel incoherent diffusive conductors
\cite{Jedema:01,Lou:07,Zaffalon:05}. One of the main results of the
present work is the observation of a \textit{gate control} of $MV$,
as a consequence of quantum interferences, which contrasts with
these previous works. Such a fact not only sheds light on the
peculiar nature of spin injection in coherent few channel
conductors, but also allows to rule out non-spin injection effects
related to stray fields for example as we will see in section IV.

\begin{figure}[!hpth]
\centering\includegraphics[height=0.95\linewidth,angle=0]{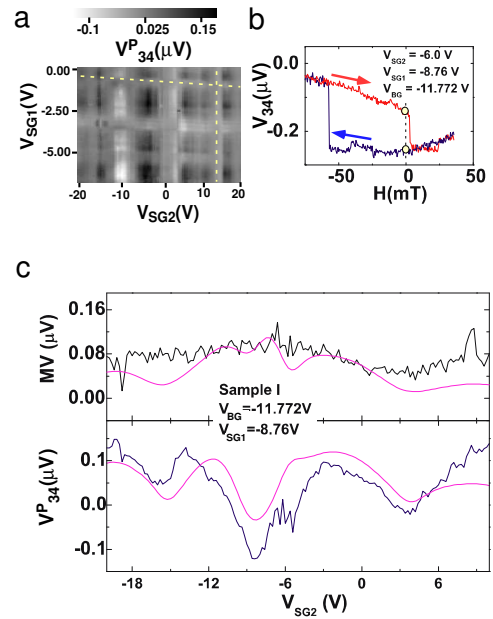}
\caption{\textbf{a.} Greyscale plot showing the "tartan" pattern of
the non local voltage $V^{P}_{34}$ of sample I in the P
configuration. \textbf{b.} "Minor hysteresis loop" for the non-local
voltage $V_{34}$ of sample I for $V_{SG1}=-8.76V$, $V_{SG2}=-6.00V$
and $V_{BG}=-11.772V$. \textbf{c.} $V^{P}_{34}$ and $MV$ as a
function of $V_{SG2}$. In purple, the prediction from the
multi-terminal scattering theory of ref. 27 with the parameters
described in the appendix.
}%
\label{greyscale}%
\end{figure}

As soon as a metallic electrode is deposited on the top of a SWNT, a
scattering region is created below the contact, which partially
decouples the two sides of the nanotube defined by the electrode.
The multi-dot nature of our devices appears on figure
\ref{greyscale}a, where $V^{P}_{34}$ is represented in a greyscale
plot as a function of $V_{SG1}$ and $V_{SG2}$ for sample I. We
observe white horizontal and vertical stripes, rather regularly
spaced, which correspond to negative anti-resonances in
$V^{P}_{34}$. Such a "tartan" pattern is very much alike the
stability diagram of a double quantum dot in the electrostatically
decoupled regime \cite{vanderWiel:03}. The stripes correspond to
discrete energy levels "engineered" by defining the 3 different
sections of the nanotube with the 4 electrodes. The fact that
horizontal as well as vertical stripes are observed shows that the
side gate electrodes control essentially independently different
parts of the device, which carry different energy levels. Our
devices can be seen as a series of three Fabry-Perot electronic
interferometers with local gate control. The nature of the coupling
between these interferometers is a crucial question for the
development of orbitally phase coherent spintronics.

One can measure the spin signals by placing the system in the
remanent state of magnetization either in the P or in the AP
configuration (for samples with a sufficiently high stability). This
is done by imposing to the device a "minor loop" which is
represented in figure \ref{greyscale}b. In such a cycle, the
magnetic field is swept in such a way as to reverse selectively one
magnetization without reversing the other. Depending on how the
external field is swept back to zero, one can reach either the P or
the AP configuration. Then, for each of these configurations, we
measure in a single shot the gate dependence of $V_{34}$. This
method has been used to obtain $MV$ in figure 2c. The existence of
quasi bound-states inside the nanotube induces variations of both
$V_{34}$ and $MV$ as a function of the gate voltages. This effect
can be observed when $V_{SG2}$ is swept, $V_{SG1}$ being kept
constant, for example (see fig. \ref{greyscale}c).  The interference
fringes observed correspond to the "tartan" pattern of figure
\ref{greyscale}a.  For $V_{SG1}=-8.76V$ and $V_{BG}=-11.772V$,
$V^{P}_{34}$ and $V^{AP}_{34}$ evolve almost in parallel as a
function of $V_{SG2}$. This results in a weakly gate dependent $MV$
with a constant positive sign as shown in figure \ref{greyscale}c.
We find that $V^{P}_{34}$ and $V^{AP}_{34}$ can be of opposite sign,
as well as of the same sign depending on the values of $V_{SG2}$ and
$V_{SG1}$. In carbon nanotubes, this phenomenon originates both from
transverse and longitudinal size quantization.

In the spectroscopy of our devices, Coulomb blockade effects are
generally absent (see e.g. Fig. \ref{nonlocalMR}b). This motivates a
comparison between our data and the non-interacting scattering model
of ref. 27 (see appendix for details). This model uses four
scattering channels, to account for the up/down spins and the K and
K' orbitals of carbon nanotubes. For simplicity, we assume that the
spin and K/K' degrees of freedom are conserved along the whole
device. Between two consecutive contacts $i\in \{1,2,3\}$ and
$j=i+1$, electrons acquire a "winding" quantum mechanical phase
$\delta _{ij}$. The effect of each metal/nanotube contact is
described with a scattering matrix which depends on the contact
transmission probability. In the case of a ferromagnetic contact, we
also take into account the spin-polarisation of the transmission
probabilities and the Spin Dependence of Interfacial Phase Shifts
\cite{Cottet:06,Cottet:06b}. This scattering model is fully
coherent, i.e. the phase of the electronic wave function is
conserved even when electrons pass in the nanotube sections below
the ferromagnetic contacts 2 and 3. The results of the scattering
theory of ref. 27, shown in magenta in figure \ref{greyscale}c, are
in qualitative agreement with our data. The variations of
$V^{P}_{34}$ are well accounted for as well as the sign and the
order of magnitude of $MV$. Importantly, the coherent model of ref.
27 involves resonance loops which are extended on several sections
of the nanotube, e.g. between leads 1 and 3.

The gate modulations of $V^{P}_{34}$ as well as the gate dependence
of $MV$ is a natural consequence of delocalization of the electronic
wave function in our devices. Similarly to optics, the multiple
reflections at the contacts give rise to (electronic) interference
which lead to gate modulations of the physical signals. It is
important to note that the origin of this gate modulation is not
related to the spin-orbit interaction which lead to energy
splittings of the order of $0.4meV$ in SWNTs\cite{Kuemmeth:08}. This
fact will become even clearer in section V where we identify the
energy scale responsible of the modulations as the single particle
level spacing of one of the NT section (namely section $12$).
Finally, it is important to note that here, contrary to the
multichannel diffusive case, coherence naturally couples the spin
and the charge of carriers. Therefore,  a non-local measurement does
not "separate spin and charge transport" as is often
stated\cite{Tombros:06} in the coherent few channel case. Rather, it
gives a new path for manipulating spin information with electric
fields at low temperature.

\section{Background magnetoresistance and stray field effects}

\begin{figure}[!h]
\centering\includegraphics[height=0.45\linewidth,angle=0]{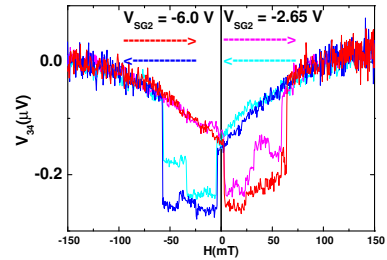}
\caption{\newline Non-local voltage $V_{34}$ for sample I as a
function of the external magnetic field $H$ for $V_{SG2}=-6.0V$ and
$-2.65V$. The curves for $V_{SG2}=-2.65V$ has been shifted up to
make them coincide with those for $V_{SG2}=-6.0V$ at zero field.
}%
\label{Background}%
\end{figure}

As one can see in figure \ref{device}b, there is a finite background
for the non-local voltage as a function of the magnetic field,
superimposed to the hysteresis. This might question the effect of
the stray fields on the observed signals. Note, however, that no
background is observed in figure \ref{nonlocal}a and c, a behavior
which is common to the majority of our samples. This makes the
device essentially insensitive to stray fields for the majority of
samples studied. In order to rule out the stray field effects for
sample I, we present in figure \ref{Background} hysteresis loops for
two of the different gate voltages, namely $V_{SG2}=-6.0V$ and
$V_{SG2}=-2.65V$. The curves for $V_{SG2}=-2.65V$ have been shifted
up to make them coincide with those for $V_{SG2}=-6.0V$ at zero
field. As one can see on this figure, while the backgrounds are
almost exactly the same (up to small gate shifts), the MV's clearly
differ. Therefore, the observed gate dependence of the MV for sample
I cannot be attributed to stray field effects.

\section{Anomalous non-local magnetoresistance}

\begin{figure}[!pth]
\centering\includegraphics[height=0.85\linewidth,angle=0]{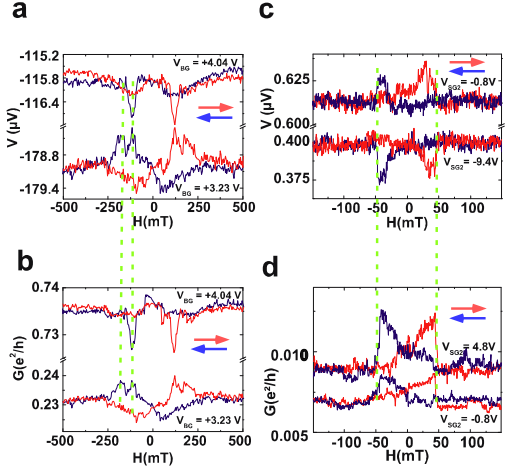}
\caption{\textbf{a.} Non-local voltage $V_{34}$ for sample II as a
function of the external magnetic field $H$ for $V_{SG1}=0.00V$,
$V_{SG2}=0.00V$ and $V_{BG}=4.04V$ or $V_{BG}=3.23V$. \textbf{b.}
Similar plot for $G$ of sample II. \textbf{c.} Non-local voltage
$V_{34}$ for sample III as a function of $H$ for $V_{SG1}=-5.00V$,
$V_{BG}=-5.455V$ and $V_{SG2}=-9.4V$ or $V_{SG2}=-0.8V$. \textbf{d.}
Similar plot for $G$ of sample III for $V_{SG1}=-5.00V$,
$V_{BG}=-5.455V$ and $V_{SG2}=4.8V$ or $V_{SG2}=-0.8V$. }
\label{nonlocal}
\end{figure}

In the multichannel diffusive incoherent regime, a hysteretic
\textit{non-local voltage} can arise, but one can show that the
\textit{intrinsic locality of charge transport} makes it very
difficult for the conductance $G=dI_{12}/dV_{sd}$ to depend on the
relative magnetic configuration of the ferromagnetic electrodes
\cite{Takahashi:03,Zutic:04}. This contrasts with our devices as
shown in fig. \ref{nonlocal}b and d where $MG \neq 0$ is obtained.
In order to show that the spin signals observed in $G$ and $V_{34}$
arise from a property of the device as a whole, it is crucial to
measure $G$ and $V_{34}$ simultaneously. In figure \ref{nonlocal}a
and b (resp. \ref{nonlocal}c and d), the magnetic field dependences
of the non-local voltage and the conductance of sample II (resp.
III) are shown for different gate voltages.  A hysteresis is
observed simultaneously for both quantities upon cycling the
magnetic field. For the measurements of figure \ref{nonlocal}a and
b, contrarily to the two other samples presented in this article,
the external magnetic field has been applied in plane,
\textit{perpendicular to the magnetic anisotropy} of the
ferromagnetic electrodes. In such a situation, the motion of the
magnetic domains often displays a complex behavior which is revealed
by the complex switching features of both $V_{34}$ and $G$ in figure
\ref{nonlocal}a and b. Because of their complexity, these features
show that the hysteretic behaviors of $V_{34}$ and $G$ have strong
correlations. As expected, we obtain more regular switchings if the
magnetic field is applied \textit{along} the easy axis anisotropy,
as shown in figures \ref{device}b, \ref{nonlocal}c and
\ref{nonlocal}d. As highlighted by the vertical dashed green lines,
both the shape and the sign of the spin signals are again strongly
correlated, which confirms that they have the same physical origin
i.e. the change in the relative magnetic configuration of the two
ferromagnetic contacts. Due to quantum interferences, $MG$ naturally
depends on $V_{SG2}$ (see figure \ref{nonlocal}b and
\ref{nonlocal}d)\cite{Cottet:09}. Therefore, we observe a spin field
effect transistor action which is non-local with respect to the
position of both the ferromagnetic electrodes and the gates. Note
that in figure \ref{nonlocal}a and \ref{nonlocal}c, we observe a
\textit{negative} $MV$. This behavior is specific to the coherent
regime and can be reproduced with the model of ref. 27.

\begin{figure}[!pth]
\centering\includegraphics[height=1.05\linewidth,angle=0]{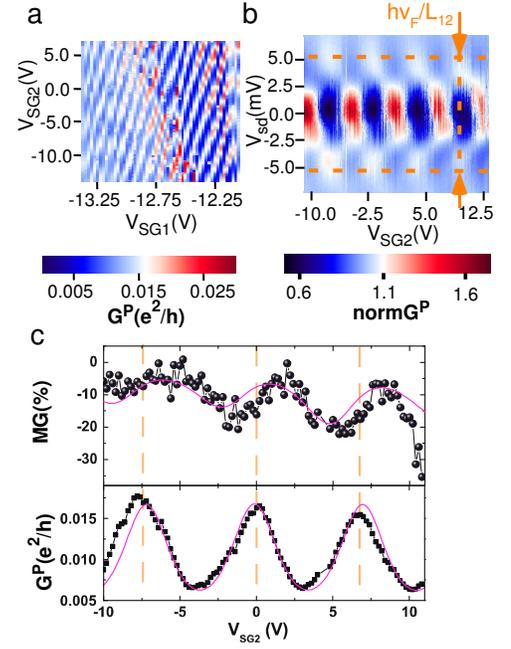}
\caption{\textbf{a.} Colorscale plot of $G^{P}$ as a function of
$V_{SG1}$ and $V_{SG2}$ for sample III. \textbf{b.} Colorscale plot
of the normalized $G$ of sample III as a function of $V_{sd}$ and
$V_{SG2}$. \textbf{c.} Simultaneous variations of $MG$ and $G^{P}$
as a function of $V_{SG2}$ for sample III.}%
\label{nonlocalMR}%
\end{figure}

The dependence of $G$ and its hysteretic part $MG=100
(1-G^{AP}/G^{P})$ on the side gate voltages further reveals how the
spin signals are affected by non-local quantum interferences. Figure
\ref{nonlocalMR}a displays the colorscale plot of $G$ as a function
of $V_{SG1}$ and $V_{SG2}$ for sample III. As indicated by the
tilted red stripes, interference fringes are observed in the
conductance. The modulations in $G$ are controlled essentially by a
single winding phase, namely $\delta_{12}$, which can be tuned via
$V_{SG2}$ or $V_{SG1}$.  As shown in figure \ref{nonlocalMR}b, the
colorscale plot of the normalized $G$ as a function of $V_{sd}$ and
$V_{SG2}$ displays the characteristic Fabry-Perot pattern with a
level spacing of about $5 meV$, consistent with the lithographically
defined length $L_{12}$ of about $300nm$ and a Fermi velocity of
$8\times 10^5 m/s$. The simultaneous measurement of $G$ and $MG$ as
a function of $V_{SG2}$ is shown in figure \ref{nonlocalMR}c. Here,
we have measured $MG$ versus $V_{SG2}$ by recording a full
hysteresis cycle for each set of gate voltages. As shown on the
bottom panel, $G^{P}$ oscillates from $0.008$ to $0.017 \times
e^2/h$ when $V_{SG2}$ is swept. Oscillations of about $30\%$ are
also found in $MG$. The solid magenta lines correspond to the result
of the scattering theory at $T=4.2K$. We find a very good agreement
with our experimental findings.  From the theoretical fit of figure
\ref{nonlocalMR}c, we conclude that $MG$ varies due to changes in
$\delta _{34}$ but also $\delta _{23}$ and $\delta _{12}$. First,
the gate electrode 2, which is nearby section 34 of the device, also
acts on $\delta _{12}$ and $\delta _{23}$ thanks to the long range
nature of Coulomb interaction ( an effect described e.g. by $\beta
_{12}\neq 0$ in the appendix). Second, the $MG$ signal is also
affected by $V_{SG2}$ due to the spatial extension of the carriers
wave function over the whole device. The non-local transistor-like
action shown in figure \ref{nonlocalMR}c is therefore non-local
electrostatically and quantum mechanically. Importantly, in figure
\ref{nonlocalMR}c, the position of the maxima in $G^{P}$ do not
coincide with those of $MG$ as highlighted by the vertical orange
dashed lines. This reveals that $G^{AP}$ oscillates in a similar
fashion as $G^{P}$ but with a \textit{different phase}. The phase
shift between $G^{P}$ and $G^{AP}$ clearly illustrates that the
phase of the carriers is conserved upon scattering below the
ferromagnetic contact 2. Indeed, this effect can only be explained
by invoking coherent electronic wave functions which extend from
contact 1 to 3 at least, and give rise to spin-dependent resonance
effects sensitive to the magnetic configuration of both leads 2 and
3. The theoretical curve of figure \ref{nonlocalMR}c reproduces
accurately this effect. We conclude that, in our devices, both the
spin and the orbital phase are conserved over the whole active part
of the nanotube, \textit{even below the ferromagnetic contacts}.

\section{Conclusion}
In this work, we have studied various non-local transport phenomena
in single wall carbon nanotubes connnected to two ferromagnetic and
two normal electrodes. These multiterminal spintronics devices
exploit actively both the spin and the orbital phase degrees of
freedom on the same footing, in spite of the use of ferromagnetic
elements. These findings could have interesting implications for the
manipulation of the electronic spin in nanoscale conductors.

\section{Appendix : Modeling of our devices}

Throughout the paper, we use the theory of reference 27 to explain
our experimental findings. Each of our device is characterized by
the set
\newline
$\{T_{1K},T_{1K'},T_{2},P_{2},\varphi^{R}_{2},\Delta
\varphi^{R}_{2},\varphi^{T}_{2}, \Delta \varphi^{T}_{2},
T_{3},P_{3},\varphi^{R}_{3}, \Delta
\varphi^{R}_{3},\\
\varphi^{T}_{3}, \Delta \varphi^{T}_{3},
T_{4K},T_{4K'},C_{Q12},C_{Q23},C_{Q34} \}$, where $T_{1(4) K[K']}$
is the transmission probability from the normal electrode $1(4)$ to
the nanotube for the K[K'] orbital, $T_{2(3)}$ is the transmission
probability between the two nanotube sections adjacent to contact
$2(3)$, $P_{2(3)}$ is the corresponding tunnel spin polarization,
$\varphi^{T[R]}_{2,(3)}$ is the spin averaged scattering phase for
an electron transmission below the contact 2(3) [an electron
reflection against 2(3)], and $\Delta \varphi^{R}_{2,(3)},\Delta
\varphi^{T}_{2,(3)}$ are the Spin Dependence of Interfacial Phase
Shifts (SDIPS) at contact 2(3) (see ref. 27). Due to the unitarity
of the contact scattering matrices, the transmission from lead 2(3)
to the nanotube is set by the above parameters. The results of the
scattering theory at $T=4.2K$ shown in magenta correspond to the
set\newline $\{0.25,0.85,0.002,0.4,\pi,0,0,0,
0.45,0.4,\pi,0,0,0,0.29,\\
0.9,11,2.2,18.0\}$ in figure \ref{greyscale}c and to the
set\newline $\{0.89,0.89,0.000035,0.8,\\
\pi, 0.3 \pi,0.295 \pi, 0.7 \pi, 0.3,0.8,\pi,0.3 \pi, 0.175 \pi,0,
0.95,\\
0.95,31,0.12,5.0\}$ in figure \ref{nonlocalMR}c. The capacitances
are in $aF$ units. For the second case, contrarily to the case of
sample I, we had to include a finite SDIPS at the ferromagnetic
contacts to enhance the amplitude of the oscillations in $MG$
\cite{Cottet:06,Cottet:06b}.

We emphasize that the above parameters are subject to several
constraints which minimize substantially the allowed phase space for
our fitting procedure. The capacitances can be estimated from the
resonance patterns in the $G$ and $V_{34}$ greyscale plots (see
section V). The transmission probabilities can be estimated from the
measurement of the two probe conductance of each section of the
device at room temperature. The  values of $G$, $MG$, $V_{34}$ and
$MV$ measured at low temperature constraint further the transmission
probabilities but also the scattering phases and the tunnel spin
polarizations. Note that for sample III, we find a very small value
of $T_2$ combined with a high value of $P_{2,3}$ and $\Delta
\varphi^{T}_{2,(3)}$. These parameters are necessary to obtain the
high $MG$ and very low $G^P$ observed in figure \ref{nonlocalMR}c.
The observed zero bias anomaly in $G$ is a possible signature of
electron-electron interactions. This effect is compensated in figure
4b by normalizing $G$ by its average $V_{sd}$-dependence over all
the gate voltages presented in the figure.

To describe the influence of the gate voltages on the circuit, we
introduce the relation $\delta _{ij}=\pi C_{Qij}(\alpha
_{ij}V_{SG1}+\beta _{ij}V_{SG2})/e$, $e$ being the elementary
charge, $C_{Qij}=2e^{2}L_{ij}/hv_{F}$ being the quantum capacitances of
each nanotube section, the dimensionless couplings $\alpha _{ij}$
and $\beta _{ij}$ being determined by the full electrostatic problem
of our devices.

\begin{figure}[!hpth]
\caption{\newline \emph{(Figure file too big-see published version for the figure)} Electrostatic diagram of our devices. We assume
here only nearest neighbor electrostatic coupling. 
}%
\label{electrostatics}%
\end{figure}

In determining the gate dependence of the theoretically expected
signals, it is important to supplement the scattering theory with a
self-consistent determination of the electrostatic potential of the
circuit. We use a coarse-grained version of the Poisson equation
which we solve self-consistently in order to determine the different
side gate actions. Our calculation proceeds along the lines of ref.
\cite{Christen:96}. We start with the full electrostatic matrix
capacitance of our devices which can be derived from the
electrostatic diagram of figure \ref{electrostatics}. We use a
nearest neighbor scheme. The total capacitance matrix $C_{TOT}$
reads :

\begin{widetext}
\begin{eqnarray}
C_{TOT}=\left(
  \begin{array}{ccccccccc}
   C_{L} & 0 & 0 & 0 & 0 & 0 & -C_{L} & 0 & 0\\
   0 & C_{R} & 0 & 0 & 0 & 0 & 0 & 0 & -C_{R}\\
   0 & 0 & 2C_{F_{L}} & 0 & 0 & 0 & -C_{F_{L}} & -C_{F_{L}} & 0\\
   0 & 0 & 0 & 2C_{F_{R}} & 0 & 0 & 0 & -C_{F_{R}} & -C_{F_{R}}\\
   0 & 0 & 0 & 0 & C_{G_{1}} & 0 & -C_{G_{1}} & 0 & 0\\
   0 & 0 & 0 & 0 & 0 & C_{G_{2}} & 0 & 0 & -C_{G_{2}}\\
   -C_{L} & 0 & -C_{F_{L}} & 0 & -C_{G_{1}} & 0 & C_{\Sigma_{1}} & -C_{m_{1}} & 0\\
   0 & 0 & -C_{F_{L}} & -C_{F_{R}} & 0 & 0 & -C_{m_{1}} & C_{\Sigma_{2}} & -C_{m_{2}}\\
   0 & -C_{R} & 0 & -C_{F_{R}} & 0 & -C_{G_{2}} & 0 & -C_{m_{2}} &
   C_{\Sigma_{3}}\\
  \end{array}
\right)
\end{eqnarray}

with $C_{\Sigma_{1}}=C_{L}+C_{F_{L}}+C_{m_{1}}+C_{G_{1}},
C_{\Sigma_{2}}=C_{m_{1}}+C_{F_{R}}+C_{F_{L}}+C_{m_{2}},
C_{\Sigma_{3}}=C_{R}+C_{F_{R}}+C_{m_{2}}+C_{G_{2}}$

\end{widetext}

In principle, we should determine self-consistently the
electrostatic potentials on each section of the nanotube using the
full scattering matrix of the problem. This would go beyond the
scope of this work. For the sake of simplicity, we assign a constant
value for the electrochemical capacitance of each section. This
assumption is reasonable in our case because the high coupling of
the SWNT to the normal electrodes reduces the energy dependence of
the electrochemical capacitance. The self-consistent equation for
the electrostatic potential of each NT section reads :

\begin{equation}
              \left\{
              \begin{array}{lll}
              (C_{\Sigma_{1}}+2 C_{Q_{12}})\delta U_{NT12}-C_{m1}\delta U_{NT23}=C_{G_{1}}\delta V_{SG1}\\
                -C_{m1}\delta U_{NT12}+(C_{\Sigma_{2}}+2 C_{Q_{23}})\delta U_{NT23}-C_{m2}\delta U_{NT34}=0\\
              -C_{m2}\delta U_{NT23}+(C_{\Sigma_{3}}+2 C_{Q_{34}})\delta U_{NT34}=C_{G_{2}}\delta V_{SG2}
                \end{array}
              \right.
\end{equation}

Finally, we get,

\begin{widetext}

\begin{eqnarray}
\left(
\begin{array}{cc}
  \alpha_{12} & \beta_{12} \\
  \alpha_{23} & \beta_{23} \\
  \alpha_{34} & \beta_{34}
\end{array}
\right)=\left(
\begin{array}{cc}
  \frac{(C_{\mu_{2}} C_{\mu_{3}}-C^{2}_{m_{2}})C_{G1}}{C_{\mu_{1}} C_{\mu_{2}} C_{\mu_{3}}-C_{\mu_{3}} C^{2}_{m_{1}}-C_{\mu_{1}} C^{2}_{m_{2}}} & \frac{C_{m_{1}}C_{m_{2}}C_{G2}}{C_{\mu_{1}} C_{\mu_{2}} C_{\mu_{3}}-C_{\mu_{3}} C^{2}_{m_{1}}-C_{\mu_{1}} C^{2}_{m_{2}}} \\
  \frac{C_{\mu_{3}} C_{m_{1}} C_{G1}}{C_{\mu_{1}} C_{\mu_{2}} C_{\mu_{3}}-C_{\mu_{3}} C^{2}_{m_{1}}-C_{\mu_{1}} C^{2}_{m_{2}}} & \frac{C_{\mu_{1}}C_{m_{2}}C_{G2}}{C_{\mu_{1}} C_{\mu_{2}} C_{\mu_{3}}-C_{\mu_{3}} C^{2}_{m_{1}}-C_{\mu_{1}} C^{2}_{m_{2}}} \\
  \frac{  C_{m_{1}}C_{m_{2}} C_{G1}}{C_{\mu_{1}} C_{\mu_{2}} C_{\mu_{3}}-C_{\mu_{3}} C^{2}_{m_{1}}-C_{\mu_{1}} C^{2}_{m_{2}}} & \frac{(C_{\mu_{2}} C_{\mu_{1}}-C^{2}_{m_{2}})C_{G2}}{C_{\mu_{1}} C_{\mu_{2}} C_{\mu_{3}}-C_{\mu_{3}} C^{2}_{m_{1}}-C_{\mu_{1}} C^{2}_{m_{2}}}
\end{array}
\right)
\end{eqnarray}

\end{widetext}

For a realistic set of capacitances $\{C_{L},C_{F_L},C_{R},C_{F_R},
C_{m_1},C_{m_2},C_{G1},C_{G2},C_{Q12},C_{Q23},C_{Q34}\}$ of about
$\{10 aF,10 aF,10aF, 10aF, 100aF,100aF,1aF,\\
1aF,30aF,1aF,10aF\}$, we find $\alpha's$ and $\beta's$ which are in
good agreement with the observed slopes in the different tartan
patterns. For example, for the above parameters, we get the
following coupling matrix :
\begin{eqnarray}
\left(
\begin{array}{cc}
  0.00837 & 0.00311 \\
  0.00441 & 0.00502 \\
  0.00328 & 0.01131
\end{array}
\right)
\end{eqnarray}
For each fitting procedure, one has to adjust the values of the
$\alpha's$ and the $\beta's$ in order to account for the gate
dependence of the observed signals. We use values which are
consistent with the above determination. Note that we have omitted
the influence of the back gate voltage here since it is set to a
constant value in our measurements.

\begin{acknowledgments}
We  are indebted with M. Aprili for discussions and sharing his thin
film deposition equipment at the beginning of this work and with A.
Thiaville, S. Rohart and J.-Y. Chauleau for illuminating discussions
on the micromagnetics of the NiPd and for MFM characterization of
our ferromagnetic contacts. The devices have been made  within the
consortium Salle Blanche Paris Centre. This work is supported by the
EU contract FP6-IST-021285-2 and by the CNano Ile de France contract
SPINMOL.

\end{acknowledgments}

\end{document}